\documentclass[aps,twocolumn,superscriptaddress]{revtex4}

\usepackage{amsmath,amssymb}
\usepackage{epsfig}
\usepackage{graphicx}
\begin{document}

\title{
Inhomogeneous post-inflationary $\Lambda$CDM cosmology as a moduli space expansion
}

\author{Toby Wiseman}
\affiliation{Theoretical Physics Group, Blackett Laboratory, Imperial College, London SW7 2AZ, U.K.}
\email{t.wiseman@imperial.ac.uk}
\author{Benjamin Withers}
\affiliation{Theoretical Physics Group, Blackett Laboratory, Imperial College, London SW7 2AZ, U.K.}
\affiliation{Department of Mathematical Sciences, Science Laboratories, South Road, Durham DH1 3LE, U.K.}
\email{b.s.withers@durham.ac.uk}
\date{October 2010}

\begin{abstract}

We model the large scale late time universe as a $\Lambda$CDM cosmology driven by cosmological constant and perfect dust fluid. Our aim is to find new solutions in the matter and $\Lambda$ epoch consistent with inflationary initial conditions, namely that to the far past in the matter era the cosmology tends to a flat FLRW solution.
We identify the moduli degrees of freedom that parametrize the flat  $\Lambda$-dust FLRW solution and then promote these moduli to slowly varying functions of the spatial coordinates and show how to solve the Einstein equations in a comoving gradient expansion, controlled by the cosmological constant length scale. 
Our initial conditions ensure that the approximation remains under control to the far past of the matter era, and to the far future of $\Lambda$ domination.
The solution is fully non-perturbative in the amplitude of the metric deformation, and we explicitly construct it to fourth order in derivatives.
A general $\Lambda$-dust universe dominated by $\Lambda$ in the future is characterized by a 3-metric and a stress tensor (with positive trace) defined on the future conformal boundary. The new cosmologies with inflationary initial conditions are characterized only by the boundary 3-metric, the stress tensor being locally determined entirely in terms of that metric. 

\end{abstract}

\pacs{}

\maketitle

\section{Introduction}

At our relatively late epoch the matter content of the universe is well approximated by a cosmological constant, $\Lambda$, and dark matter \cite{Riess:1998cb,Perlmutter:1998np}. We live at an interesting time, approximately at $\Lambda$-matter equality, before which stretches a very long matter era, and to our future an infinite period of $\Lambda$ domination. In this paper, as in our previous work  \cite{Wiseman:2010kp}, we take the view that it is interesting to focus on solutions to the Einstein equations which describe the late time dynamics of our universe. We wish to use the natural late time scale, set by the cosmological constant, in order to separate the features of our cosmology that  are relevant at late times from those that are irrelevant. We will model the $\Lambda$-matter era of a $\Lambda$CDM universe on large scales as a cosmology with cosmological constant and perfect dust fluid. Our aim is to find new tools to characterize and construct the most general inhomogeneous anistropic $\Lambda$-dust cosmologies relevant for describing our late time universe throughout the $\Lambda$ and matter eras. 

Taking only $\Lambda$ and dust matter gives an expanding cosmology which inevitably has a dust dominated big bang to the past. For a $\Lambda$-dust solution to approximate our universe we must impose that the behaviour in the early dust era is consistent with inflationary initial conditions, namely that going back into the far past of the dust era the universe is increasingly homogeneous, isotropic and spatially flat. Said another way, we wish to construct $\Lambda$-dust cosmologies which tend to a flat FLRW solution in the far past of the dust era, and differ from it only by a \emph{growing} deformation. We term such a solution a \emph{post-inflationary $\Lambda$-dust} cosmology. We note that for a flat $\Lambda$-dust FLRW cosmology the only length scale is given by the cosmological constant, and we will indeed see that the presence of $\Lambda$ will be crucial to our construction of solutions.

Cosmological perturbation theory \cite{Lifshitz:1945du,Lifshitz:1963ps,Bardeen:1980kt,Kodama:1985bj} provides the conventional tool by which one constructs such deformations of flat FLRW.
Despite the success of perturbation theory we believe it is still interesting to pursue a different construction of inflationary $\Lambda$CDM cosmologies. 
In this paper, we detail a new construction which instead takes the form of a derivative expansion. The construction can be regarded as a moduli space approximation, where the natural moduli of the flat FLRW $\Lambda$-dust cosmology are identified, and then allowed to have slow spatial comoving variations controlled by the cosmological constant length scale. 
In cosmological perturbation theory, taking growing initial perturbations is crucial in order to consistently describe early times within the perturbative approximation. Similarly in our approach, this initial condition is again important in allowing control to be retained in the early matter era within the gradient expansion approximation. Whereas at late times in perturbation theory, growing perturbations may grow to the point of non-linearity, in our gradient expansion, the construction of the solution is fully non-perturbative in the metric deformation from FLRW and therefore this does not present a problem.

Previous works have utilised the Hamilton-Jacobi framework to construct derivatively expanded cosmologies in the context of inflation \cite{PhysRevD.42.3936,Salopek:1990re,Parry:1993mw}. The work presented here differs from these approaches in two ways. Firstly, we are constructing solutions for the duration of $\Lambda$ and matter dominance, rather than an inflationary epoch. Secondly and most importantly, we impose physical inflationary initial conditions.

In our previous work  \cite{Wiseman:2010kp}  we have discussed how the general $\Lambda$CDM cosmology that is $\Lambda$ dominated at late times can be characterized by its late time asymptotic behaviour, in particular the future conformal boundary 3-metric and stress tensor. We find that the cosmologies we construct here are indeed $\Lambda$ dominated at late times, and are characterized precisely by their future conformal boundary 3-metric, with their stress tensor being fully, and perhaps surprisingly, locally determined in terms of this metric. 

The construction we use is analogous to the AdS-CFT discussion of emergent hydrodynamics from deformations of the homogeneous black brane \cite{Bhattacharyya:2008jc,Bhattacharyya:2008xc}. There, the bulk solutions are constructed as a derivative expansion controlled by the AdS length scale, and are shown to be characterized by a boundary stress tensor that describes a hydrodynamic theory. This derivative expansion may be thought of as a moduli space approximation, where the moduli give the physical degrees of freedom of the homogeneous black brane. The key insight in that setting is understanding the boundary conditions, namely that the asymptotic boundary geometry is unperturbed and requiring the future horizon remains regular. In our cosmological setting we may regard the boundary condition that at early times the deformation from FLRW contains no decaying component as being the analog of the horizon regularity condition. However, an important difference is that in our setting the boundary metric must be deformed in order to obtain interesting dynamics.

The structure of this paper is as follows. In the next two sections we detail the Einstein equations and briefly review our previous work on characterizing general $\Lambda$ dominated $\Lambda$-dust cosmologies from their late time behaviour. We then discuss the nature of inflationary initial conditions, and review these explicitly using cosmological perturbation theory. In section \ref{sec:moduli} we detail the moduli space approximation for deformations of flat FLRW $\Lambda$-dust cosmologies, show how to impose inflationary initial conditions, and compute it explicitly up to fourth order in comoving spatial derivatives. In section \ref{sec:stress} we discuss the form of the boundary stress tensor that characterizes the solutions. Finally in section \ref{sec:smoothing} we briefly describe the use of Ricci flow to provide a natural smoothing flow in the space of post-inflationary $\Lambda$-dust cosmologies, which gives flat FLRW as a stable fixed point.

\section{The Einstein equations}

We will take the matter content to be cosmological constant $\Lambda$ and cold dark matter.  We choose units so $8 \pi G_N = 1$. Since we will later be working in a derivative expansion, controlled by the scale of the cosmological constant, we will be interested in large scales in the universe where we expect the dark matter to be well modeled  by pressureless perfect  dust fluid. To begin we write the spacetime metric in a Gaussian normal form,
\begin{eqnarray}
\label{eq:metric1}
ds^2 =  \frac{1}{y^2} \left( - \frac{dy^2}{H^2}  + a^2(y) g_{ij}(x,y) dx^i dx^j \right)
\end{eqnarray}
where $x^i$ for $i=1,2,3$ give spatial coordinates on a constant $y$ time surface and $H^2 = \Lambda/3$, and would give the Hubble rate for the de Sitter geometry corresponding to the cosmological constant $\Lambda$. 
Our time coordinate $y$ is related to the usual proper time $t$ as $y \sim e^{- H t}$, so that $y \rightarrow 0$ is the far future, and increasingly positive $y$ corresponds  to the further back in the past. This choice of time coordinate simplifies discussion of the late time behaviour of our cosmologies, and conveniently compactifies future infinity to the conformal boundary $y = 0$ provided $g_{ij}$ is regular there.

For our normal coordinates the vector $\partial/\partial y$ is tangent to timelike geodesics. Hence we have the freedom to choose these constant $y$ surfaces to comove with the dust fluid. Then the stress tensor takes the form,
\begin{eqnarray}
T_{yy} = \frac{1}{H^2 y^2} \left( \Lambda + \rho \right) \; , \quad T_{yi} = 0 \; , \quad T_{ij} =  - \frac{\Lambda}{y^2} a^2 g_{ij}\nonumber
\end{eqnarray}
and we do not have to introduce a velocity field for the dust, which is an important technical simplification in what follows. In particular it will allow us to straightforwardly identify the appropriate moduli of the flat FLRW solution.

We may write the flat FLRW solution by taking $g_{ij}(y,x) =  \delta_{ij}$ with the scale factor $a(y)$ obeying, 
\begin{eqnarray}
\frac{\ddot{a}}{a} - \frac{2}{y} \frac{\dot{a}}{a} + \frac{1}{2} \left( \frac{\dot{a} }{a} \right)^2 = 0 \; , \quad \frac{\ddot{a}}{a} - \frac{1}{y} \frac{\dot{a}}{a} = - \frac{1}{6 H^2 y^2} \rho_{\text{\tiny{FLRW}}}\nonumber
\end{eqnarray}
where $\dot{} = \partial / \partial y$ and $ \rho_{\text{\tiny{FLRW}}}$ denotes the spatially homogeneous isotropic dust density. The
solution is $a^2(y) = \left( 1 - \tfrac{\rho_0}{12 H^2} y^3 \right)^{4/3}$, with $\rho_0$ a constant of integration
so that $ \rho_{\text{\tiny{FLRW}}} \sim \rho_0 y^3 + O(y^4)$ asymptotically at late times, $y \rightarrow 0$. However, by the trivial global scaling $y \rightarrow \lambda y$, $x^i \rightarrow \lambda x^i$, we may bring the scale factor to the form,
\begin{eqnarray}
a^2(y) = (1 - y^3)^{4/3}\label{FLRWScaleFactor}
\end{eqnarray}
without loss of generality, and from now on we take $a(y)$ to be defined in this way.

We now perform a $1+3$ decomposition of the Einstein equations. The Einstein equations then give a scalar and a vector equation with respect to the 3-metric $g_{ij}$,
\begin{eqnarray}
 \ddot{g} + \left(- \frac{1}{y} + \frac{\dot{a}}{a} \right) \dot{g} - \frac{1}{2} \dot{g}_{ij} \dot{g}^{ij} + 6 \left( \frac{\ddot{a}}{a} - \frac{1}{y} \frac{\dot{a}}{a} \right) &=& - \frac{1}{H^2 y^2} \rho \nonumber \\
\nabla^j \dot{g}_{ij} - \nabla_i \dot{g} &=& 0\nonumber
\end{eqnarray}
together with the tensor equation,
\begin{eqnarray}
 \ddot{g}_{ij} + \left( - \frac{2}{y} + 3 \frac{ \dot{a}}{a} \right) \dot{g}_{ij}  + \frac{1}{8} g_{ij} \left( \dot{g}_{mn}\dot{g}^{mn}- \dot{g}^2 \right)    && \nonumber \\
-  \dot{g}_{im} \dot{g}_{j}^{~m} + \frac{1}{2} \dot{g} \, \dot{g}_{ij} + \frac{2}{H^2a^2} \left( R_{ij} - \frac{1}{4} g_{ij} R \right) &=& 0 .
 \label{eq:tensoreq}
\end{eqnarray}
Indices are raised and lowered with respect to $g_{ij}$, $\nabla_i$ is the Levi-Civita connection of $g_{ij}$, $R_{ij}$ is its Ricci tensor, and we define $\dot{g}_{ij} = \partial_y g_{ij}$, $\dot{g} = g^{ij} \dot{g}_{ij}$ and likewise $\ddot{g}_{ij} = \partial^2_y g_{ij}$, $\ddot{g} = g^{ij} \ddot{g}_{ij}$. The dust equation of motion is implied by these Einstein equations due to the contracted Bianchi identity.

The scalar equation simply determines the dust fluid density. The vector equation is non-trivial. Let us denote the vector,
\begin{eqnarray}
\Phi_i \equiv {\nabla}^j \dot{g}_{ij} -{\nabla}_i \dot{g}\nonumber
\end{eqnarray}
so that the equation is $\Phi_i = 0$. The tensor equation \eqref{eq:tensoreq} implies,
\begin{eqnarray}
\label{eq:dconst}
\partial_y \Phi_i + \left( - \frac{2}{y} + 3 \frac{ \dot{a}}{a} \right) \Phi_i =  - \frac{1}{2} \dot{g} \, \Phi_i
\end{eqnarray}
and hence, as we would expect for a 1+3 decomposition, provided this constraint is satisfied on some constant $y$ slice, it will always be satisfied. Hence we may regard it as constraining the space of solutions.

\section{Review of late time expansion}

In this section we briefly review the late time characterization of $\Lambda$-dust cosmologies that become dominated by $\Lambda$ \cite{Wiseman:2010kp}. For such solutions the asymptotic behaviour was deduced by Starobinsky in the context of inflation \cite{Starobinsky:1982mr}, and studied carefully more recently in \cite{Rendall:2003ks,Rodnianski:2009de}. 
At late times the general solution which is $\Lambda$ dominated goes as,
\begin{eqnarray}
\label{eq:soln}
a(y)^2 g_{ij}(y,x) &=& \bar{g}_{ij} + \frac{y^2}{H^2} \left( \bar{R}_{ij} - \frac{1}{4} \bar{g}_{ij} \bar{R} \right) + y^3 \bar{h}_{ij}  \nonumber \\
&& \quad + y^4 \bar{a}^{(1)}_{ij}+ y^5 \bar{a}^{(2)}_{ij} + O(y^6),
\end{eqnarray}
where $\bar{g}_{ij}(x)$ is a freely specifiable non-degenerate Euclidean signature 3-metric which gives the geometry of the future conformal boundary $y = 0$. All tensors in the above expression are taken to live on the 3-geometry defined by $\bar{g}_{ij}$, and indices are raised/lowered and derivatives taken with respect to this boundary metric, and $\bar{R}_{ij}$ is the Ricci tensor of $\bar{g}_{ij}$. The tensor $\bar{h}_{ij}(x)$ is also general up to certain conditions. It is symmetric and an analysis of \eqref{eq:dconst} shows it must obey the constraint, $\bar{ \nabla}^j \bar{h}_{ij} - \bar{\nabla}_i \bar{h}= 0$. The dust density goes as, $\rho = - 3H^2 \bar{h}\, y^3 + O(y^4)$, and hence we require that the trace $\bar{h}(x) < 0$ everywhere for physical dust densities. The tensors $\bar{a}^{(1)}_{ij}$ and $\bar{a}^{(2)}_{ij}$ are given in \cite{Wiseman:2010kp}. 

The choice of 3-metric $\bar{g}_{ij}$ and symmetric tensor $\bar{h}_{ij}$, with negative trace and obeying $\bar{ \nabla}^j \bar{h}_{ij} - \bar{\nabla}_i \bar{h}= 0$ then fully determine the solution at late times. All higher terms $\bar{a}^{(n)}_{ij}(x)$ are tensors determined covariantly in terms of this data $(\bar{g},\bar{h})$.  Thus we may take this pair $(\bar{g},\bar{h})$ to characterize the $\Lambda$ and matter eras of all $\Lambda$CDM cosmologies that are asymptotically $\Lambda$ dominated in the far future. The cosmologies we construct later are indeed in this class.

From the tensor $\bar{h}_{ij}$ we may simply define $\bar{T}_{ij} \equiv \bar{h}_{ij} - \bar{g}_{ij} \bar{h}$, so that $\bar{T}_{ij}$ is conserved in the sense of a stress tensor $\bar{\nabla}^j \bar{T}_{ij} = 0$. The positive dust density condition requires that its trace is positive, $\bar{T} \ge 0$. The trace is zero in the case of vanishing dust density $\rho = 0$. 

In the context of dS/CFT and cosmological holography \cite{Strominger:2001pn,Strominger:2001gp,Larsen:2002et,Larsen:2003pf,Larsen:2004kf,vanderSchaar:2003sz,McFadden:2009fg,McFadden:2010na}, a naive analysis would suggest that $\bar{T}_{ij}$ should play the role of the dual holographic stress tensor. Indeed setting $\rho=0$ so that the only matter is the cosmological constant, one can obtain a holographic renormalization group from analytic continuation of the AdS-CFT result \cite{Skenderis:2002wp, Gibbons}, and the traceless $\bar{T}_{ij}$ would indeed give the dual stress tensor. Since dust leads to a non-vanishing trace of the stress tensor, if a holographic dual theory did exist for $\Lambda$CDM one would not expect it to be a conformal field theory.

\section{Inflationary initial conditions}
\label{sec:linear}

The data $( \bar{g}, \bar{h} )$ characterize the most general $\Lambda$-dust cosmology that is $\Lambda$ dominated. However physically we are interested in $\Lambda$-dust cosmologies that have arisen from inflationary initial conditions, solutions that are deformations of flat FLRW by only growing mode perturbations in the early matter era. 
Of course in our actual universe we believe the matter era is preceded by radiation and inflation, and we would expect some decaying perturbations in the early matter epoch. However, from the perspective of a late time observer basing physics around the current cosmological constant scale, and characterizing the solution in terms of late time data $( \bar{g}, \bar{h} )$, the matter era appears to be extremely long, unnaturally so from the perspective of the late time cosmological constant scale,
and the early matter era appears to be very well approximated by a matter dominated big bang with only growing perturbations. 
The amount of decaying mode perturbation to be added in due to the fact that really there is no matter big bang, but a smooth transition to an earlier radiation epoch is vanishingly small when thought about in terms of the late time data, being suppressed parametrically by powers of the $\Lambda$-matter equality scale to the matter-radiation equality scale. It is this parametric separation of scales that controls the decoupling of relevant dynamics in the $\Lambda$-matter era of our universe from that which is irrelevant, such as the details of the inflation and radiation era.

Generic late time data $( \bar{g}, \bar{h} )$ will correspond to a universe which at early times is far from FLRW, in the sense that it will contain `decaying' deformations that become stronger as one goes further back into the early matter epoch. The task of this paper is therefore to understand how to characterize and construct the subset of cosmologies which are described by post-inflationary or growing deformations of FLRW such that the deformations become smaller as one goes to earlier and earlier times in the matter era. We will give a construction that is non-perturbative in the metric deformation from flat FLRW and instead takes the form of a derivative expansion. 
However, we have found it helpful to understand the conventional cosmological perturbation theory description of such post-inflationary $\Lambda$-dust cosmologies and this is reviewed and developed in the rest of this section.

For small metric perturbations about flat FLRW, our metric choice \eqref{eq:metric1} describes the synchronous gauge, where the residual gauge invariance has been fixed by requiring constant $y$ slices to comove with the dust. Following the conventional approach \cite{Bardeen:1980kt,Kodama:1985bj} we consider scalar, vector and tensor perturbations with respect to the spatial metric. The perturbation is then written,
\begin{eqnarray}
g_{ij}(x,y) = \delta_{ij} + \delta g^S_{ij}(x,y) + \delta g^V_{ij}(x,y) + \delta g^T_{ij}(x,y) .\nonumber
\end{eqnarray}
Let us define the differential operator,
\begin{eqnarray}
D_y \equiv \partial_y +  \left( - \frac{2}{y} + 3 \frac{\dot{a}}{a} \right) \nonumber
\end{eqnarray}
then the tensor perturbation, $\delta g^T_{ij}$, obeys the transverse traceless conditions $\partial^i \delta g^T_{ij} = \delta g^T = 0$, with indices raised and lowered with respect to the Euclidean metric $\delta_{ij}$, and the evolution equation,
\begin{eqnarray}
D_y \partial_y {\delta{g}}^T_{ij} = \frac{1}{a^2(y) H^2} \partial_k^2 \delta{g}^T_{ij} .\nonumber
\end{eqnarray}
The scalar part is decomposed as $\delta g^S_{ij}(x,y) = A(x,y) \delta_{ij} + \partial_i \partial_j B(x,y)$, where the linearized Einstein equations imply,
\begin{eqnarray}
A(x,y) &=& \bar{\phi}(x)  \nonumber \\
D_y \partial_y {B}   &=&  \frac{1}{a^2(y) H^2} \bar{\phi}(x)\nonumber
\end{eqnarray}
where $\bar{\phi}(x)$ is an arbitrary function, but only of the spatial coordinates. The vector sector $\delta g^V_{ij}$ is dynamically trivial, leading only to a $y$ independent infinitesimal diffeomorphism, and we will ignore it here.

In the flat FLRW solution \eqref{FLRWScaleFactor} the big bang occurs at $y=1$. Defining the new time, $\epsilon = 1-y$, we may describe the early matter era by considering small $\epsilon$, where $\epsilon$ behaves as a proper time coordinate. Then $a \sim \epsilon^{2/3}$, and the tensor and scalar modes behave as,
\begin{eqnarray}
\mathrm{Decaying:} && B \sim \frac{1}{\epsilon} \qquad {\delta g}^T_{ij} \sim \frac{1}{\epsilon} \nonumber \\
\mathrm{Growing:} && B \sim \mathrm{const}  \qquad {\delta g}^T_{ij} \sim \mathrm{const} .\nonumber
\end{eqnarray}
The scalar perturbation leads to a perturbation in the dust density, so $\rho(y,x) = \rho^{(0)}(y) + \delta \rho(y,x)$ where $\rho^{(0)}(y)$ is the FLRW dust behaviour, and the perturbation $\delta \rho$ behaves as
\begin{eqnarray}
\mathrm{Decaying:} && \frac{ \delta \rho }{\rho^{(0)}} \sim  \frac{1}{\epsilon} \nonumber \\
\mathrm{Growing:} && \frac{ \delta \rho }{\rho^{(0)}} \sim \epsilon^{2/3} \nonumber
\end{eqnarray}
for the scalar decaying and growing modes.

We may analytically solve the scalar perturbation for all $y$ which behaves at early times as the growing mode above;
\begin{eqnarray}
\label{eq:scalarPTsoln}
B(x) &=& c(x) - \frac{\bar{\phi}(x) }{2 H^2} \Bigg( \frac{y^2  \left( 2 + (1 + y^3)\; {}_2F_1\left[ 1,\frac{4}{3};\frac{5}{3}; y^3 \right] \right)}{3 (1 - y^3)^{1/3}} \nonumber \\
&& \qquad \qquad \qquad \qquad - \frac{3 j y^3}{1 - y^3}   \Bigg)
\end{eqnarray}
where ${}_2F_1$ is a hypergeometric function, and $j$ is the constant,
\begin{eqnarray}
\label{eq:constant}
j = \frac{2^{8/3}\sqrt{\pi}}{3^{5/2}} \frac{\Gamma[5/6]}{\Gamma[1/3]} \simeq 0.304\ldots
\end{eqnarray}
and $c(x)$ is a constant of integration and we have picked a convenient normalisation for the mode. As for the vector perturbations, the constant $c(x)$ simply leads to a $y$ independent infinitesimal diffeomorphism, and hence we will set $c(x) = 0$.

However, for the tensor mode we know of no analytic solution for the combined $\Lambda$ and matter eras and hence cannot analytically match this late time behaviour to the early growing behaviour above. However, performing a Fourier decomposition $\delta g^T_{ij}(y,x) = \int dk \, a_{ij}(k_l) u(y, k^2) e^{i H k_m x^m}$, we may numerically solve a mode with wavenumber $k_i$ which obeys,
\begin{eqnarray}\label{uequation}
D_y \partial_y {u} = - \frac{k^2}{a^2(y)} u .
\end{eqnarray}

We now consider the late time behaviour of perturbations  that grow in the early matter era. 
Firstly consider our growing scalar mode above. At late times this mode behaves as,
\begin{eqnarray}
B =  - \frac{1}{2 H^2} y^2 \bar{\phi} + \frac{3 j}{2 H^2} y^3 \bar{\phi} + O(y^4).\nonumber
\end{eqnarray}
The tensor modes are more complicated to treat and we can only numerically match the early time growing behaviour since we lack an analytic solution. In terms of the Fourier decomposition, at late times one finds,
\begin{eqnarray}
u(y, k^2) = 1 + y^2 \frac{k^2}{2} + y^3 \,b(k^2) + O(y^4).\nonumber
\end{eqnarray}
We may determine the leading long wavelength behaviour for $b(z)$ analytically by solving \eqref{uequation} order by order in $z$,
\begin{eqnarray}
b(z) =  -\frac{3j}{2}z + c_2 z^2 + O(z^3)\nonumber
\end{eqnarray}
whilst we numerically determine $c_2$ as,
\begin{equation}\label{eqc2}
c_2\simeq0.09.
\end{equation}
We may then rewrite this in position space as,
\begin{eqnarray}
&& \delta g^T_{ij}(y,x)  =\nonumber\\
&& \quad\left(  1- y^2 \frac{\partial^2}{2 H^2} + y^3 b(- \frac{\partial^2}{H^2}) +  O(y^4) \right) \bar{\Phi}_{ij}(x), \nonumber
\end{eqnarray}
for some transverse traceless data $\bar{\Phi}_{ij}(x)$.

We see that the scalar and tensor modes combine to describe a perturbation to the boundary metric, so that,
\begin{eqnarray}
\bar{g}_{ij}(x) &=& \delta_{ij} + \left( \bar{\phi} \, \delta_{ij}  + \bar{\Phi}_{ij} \right).\nonumber
\end{eqnarray}
In fact, had we included the vector perturbation and the scalar field $c(x)$ in \eqref{eq:scalarPTsoln}, we would have also obtained the contribution $\partial_{(i }\bar{\xi}_{j)}$ for some vector field $\xi_i(x)$, which is simply an  infinitesimal diffeomorphism.
We observe from perturbation theory that the data $\bar{h}_{ij}$ is then completely determined by the perturbation to $\bar{g}$ as,
\begin{eqnarray}
\bar{h}_{ij}(x) &=& - \frac{4}{3} \left( \delta_{ij} + \left( \bar{\phi} \, \delta_{ij}  + \bar{\Phi}_{ij} \right) \right)  \nonumber \\
&& + \frac{3 j}{2 H^2} \partial_i \partial_j \bar{\phi} + b(- \frac{\partial^2}{H^2}) \bar{\Phi}_{ij} .\nonumber
\end{eqnarray}

We now make two observations. Firstly we note that the general perturbation that grows in the early matter era can be characterized by the perturbation of $\bar{g}$, with the perturbation of $\bar{h}$ being completely determined by this. In a sense this is to be expected, since every perturbation wavenumber comes in both a growing and decaying time dependence at early times, we expect by choosing one behaviour we halve the data required to characterize the perturbation. 

Secondly, we observe that if we truncate $\bar{h}_{ij}$ to two derivative order we obtain, to linear order in the perturbations,
\begin{eqnarray}
\delta \bar{h}_{ij}(x) &=& - \frac{4}{3} \delta \bar{g}_{ij} + \frac{3 j}{2 H^2} \left( \partial_i \partial_j \bar{\phi} + \partial^2\bar{\Phi}_{ij} \right) + O(\partial^4) \nonumber \\
& = & - \frac{4}{3} \delta \bar{g}_{ij} - \frac{3 j}{H^2} \left( \delta\bar{R}_{ij} - \frac{1}{4} \delta\bar{R} \, \delta_{ij} \right) + O(\partial^4)\nonumber
\end{eqnarray}
where $\delta \bar{g}_{ij} = \bar{\phi} \, \delta_{ij}  + \bar{\Phi}_{ij}$ is the perturbation to the metric $\bar{g}_{ij}$, and $\delta\bar{R}_{ij}$ is the perturbation to its Ricci tensor $\bar{R}_{ij}$. 

Based on these observations, we might conjecture that the general solution which asymptotes in the far matter past to FLRW, as required for a post-inflationary universe, is characterized by a general 3-metric $\bar{g}$, with $\bar{h}$ being completely and covariantly determined by $\bar{g}$ as a derivative expansion with leading term,
\begin{eqnarray}
\label{eq:predict}
\bar{h}_{ij}(x) &=& - \frac{4}{3}  \bar{g}_{ij} - \frac{3 j}{H^2} \left( \bar{R}_{ij} - \frac{1}{4}\bar{R} \,\bar{g}_{ij} \right) \nonumber\\
&& + O( \bar{\nabla}^2 \bar{R}, \bar{R}^2 ).
\end{eqnarray}

In the next section we show how to view the boundary metric as giving the moduli for the FLRW solution, and then allowing the metric to spatially vary gives a moduli space approximation for the general inhomogeneous anisotropic $\Lambda$-dust cosmology. We will then confirm that the prediction for the boundary $\bar{h}_{ij}$ above is correct.

\section{Moduli space approximation}
\label{sec:moduli}

Let us now exhibit the moduli parameters of the flat FLRW $\Lambda$-dust solution. Firstly we perform an arbitrary time independent global coordinate transform of the spatial coordinates, so that, 
\begin{eqnarray}
\delta_{ij} dx^i dx^j \rightarrow \bar{g}_{ij} dx^i dx^j\nonumber
\end{eqnarray} 
where $\bar{g}_{ij}$ is a symmetric, non-degenerate tensor which is constant in time and space. Any such tensor $\bar{g}_{ij}$ can be obtained by a suitable choice of coordinate transform. The flat FLRW solution is then given by, $g_{ij}(y,x) = \bar{g}_{ij}$. Our linear theory discussion above shows that the inflationary cosmologies should precisely be characterized by the boundary metric $\bar{g}_{ij}$. Hence it suggests the constant components of the tensor $\bar{g}_{ij}$ here are in fact the moduli of the flat FLRW solution, and that we can consider them to vary slowly spatially in order to develop a moduli space approximation. 

We know that a general late time $\Lambda$ dominated solution for $\Lambda$-dust matter is characterized in terms of two tensors: a boundary metric $\bar{g}$ and a symmetric tensor $\bar{h}$ which obeys a conservation equation. Since our moduli appear to be precisely the boundary metric, we clearly must require some boundary conditions in order to dynamically determine $\bar{h}$. Following our discussion above, we take the boundary condition that the solution is asymptotically flat FLRW to the far past, which can be thought of as the late time physical requirement that the universe had an inflationary origin. In practice we shall see shortly that this corresponds to a growing mode boundary condition on a linear differential operator, in analogy with the cosmological perturbation theory calculation.

Let us characterize the spatial variation of $\bar{g}$ by taking the radius of curvature of the 3-metric $\bar{g}$ to be characterized by the length $L$, so that,
\begin{eqnarray}
\bar{\nabla}_{a_1} \ldots \bar{\nabla}_{a_m} \bar{R}_{ij} \simeq O( \frac{1}{L^{m+2}} ).\nonumber
\end{eqnarray}
Let us define a perturbation parameter $\lambda = 1 / (H L)^2$ in which to perform our derivative expansion. We claim for varying moduli, the solution takes the form,
\begin{eqnarray}
g_{ij}(y,x) = \bar{g}_{ij}(x) + g^{(1)}_{ij}(x,y) + g^{(2)}_{ij}(x,y) + O(\lambda^6) \nonumber
\end{eqnarray}
where $g^{(n)}_{ij}(x,y) \sim O(\lambda^{2 n})$ is a tensor, living on the geometry defined by $\bar{g}_{ij}$ and being a function of $y$, and being determined in terms of the moduli data $\bar{g}_{ij}$ by a covariant expression in $\bar{g}_{ij}$ involving $(2 n)$ spatial derivatives. We note that the derivative expansion is constrained to have only even powers by spatial parity symmetry.

An important point is that the decomposition of the metric into a leading part, $\bar{g}_{ij}$, that depends only on $x$ and subleading parts depending on both $x$ and $y$ is not unique. We will find it convenient to define this splitting by requiring that $g^{(n)}_{ij}(x,0) = 0$ for all $1 \le n$. This implies that the conformal boundary metric is simply given by $\bar{g}_{ij}$, and is not corrected by higher orders in the derivative expansion.

We will now explicitly give the tensor equation and its solution for the trivial zeroth order, and the orders determining $g^{(1,2)}_{ij}$ in the expansion. We then comment on the structure of higher orders. Having determined the solutions $g^{(1,2)}_{ij}$, we will then compute the derivative expansion of the dust density,
\begin{eqnarray}
\rho(y,x) = \rho^{(0)}(y) + \rho^{(1)}(y,x)+  \rho^{(2)}(y,x) + O(\lambda^6) \nonumber
\end{eqnarray}
where again $\rho^{(n)}_{ij}(x,y) \sim O(\lambda^{2 n})$ and depends on the solutions $g^{(m)}_{ij}$ for $m \le n$. In addition, we must also confirm that the constraint equation is satisfied order by order, and it is convenient to expand the constraint $\Phi_i$ as,
\begin{eqnarray}
\Phi_i(y,x) = \lambda \left( \Phi_i^{(1)}(y,x) + \Phi_i^{(2)}(y,x) + O(\lambda^6)\right) \nonumber
\end{eqnarray}
where the constraint $\Phi_i^{(n)}(y,x) \sim O( \lambda^{2 n} )$ and depends on the solution $g^{(m)}_{ij}$ for $m \le n$.

\subsection{Solution at trivial order:  $O( \lambda^0 )$}

Decomposing the equation \eqref{eq:tensoreq} into orders in $\lambda$ one obtains at the trivial zeroth order, $O(\lambda^0)$, simply the Friedmann equation for $a(y)$. The constraint equation is trivial at this order too. The dust density from the scalar equation is simply given by,
 \begin{eqnarray}
 \rho^{(0)}(y) & = & - 6 H^2 y^2 \left( \frac{\ddot{a}}{a} - \frac{1}{y} \frac{\dot{a}}{a} \right) =  \frac{12 H^2 y^3}{( 1 - y^3 )^2}. \nonumber
\end{eqnarray}
We note that this is just the usual FLRW dust density time dependence. One might think that this density should vary spatially in the moduli approximation, with large magnitude variations being allowed, which would seem to be at odds with the fact that the leading order behaviour is just that of the undeformed FLRW solution. However, we emphasize that we have chosen a co-moving coordinate system with the dust, and observe that on a comoving slice the dust density at leading order remains constant. This, of course, does not mean that the dust density is only perturbed in a small manner from FLRW since the constant $y$ slice, with induced metric $a^2(y) \bar{g}_{ij}(x)/y^2$, can be strongly deformed relative to the FLRW one, which simply has a Euclidean geometry with induced metric $a^2(y) \delta_{ij} / y^2$.

\subsection{Solution at two derivative order: $O( \lambda^2 )$}

The first non-trivial order is at two derivatives, $O(\lambda^2)$, where one finds the tensor equation yields,
\begin{eqnarray}
D_y \partial_y {g}^{(1)}_{ij} = - \frac{2}{a^2 H^2} \left( \bar{R}_{ij} - \frac{1}{4} \bar{g}_{ij} \bar{R} \right) \nonumber
\end{eqnarray}
where the right hand side is the Schouten tensor of the boundary metric.
We note that due to the simple tensor structure, this is really a first order \emph{ordinary} differential equation for $g^{(1)}_{ij}(y,x)$ and can by simply solved by quadrature. Defining the source functions $f^{(1)}(y) =  - \frac{2}{a^2}$ and $\bar{S}^{(1)}_{ij}(x) = \frac{1}{H^2} \left( \bar{R}_{ij} - \frac{1}{4} \bar{g}_{ij} \bar{R} \right)$ we recast the first order equation simply as,
\begin{eqnarray}
D_y \partial_y {g}^{(1)}_{ij}(y,x) = f^{(1)}(y) \bar{S}^{(1)}_{ij}(x). \nonumber
\end{eqnarray}
We note that these equations are simply tensor versions of the ones that occur for the scalar perturbations discussed before. 
A particular solution, ${g}^{(1,p)}_{ij}$ can be obtained by quadrature,
\begin{eqnarray}
{g}^{(1,p)}_{ij}(y,x) = \bar{S}^{(1)}_{ij}(x)   \frac{y^2  \left( 2 + (1 + y^3)\; {}_2F_1\left[ 1,\frac{4}{3};\frac{5}{3}; y^3 \right] \right)}{3 (1 - y^3)^{1/3}} \nonumber
\end{eqnarray}
where ${}_2F_1$ is a hypergeometric function. This particular solution
at early times behaves as,
\begin{eqnarray}
{g}^{(1,p)}_{ij}(y,x) = \bar{S}^{(1)}_{ij}(x)  \left(  \frac{j}{\epsilon} - \frac{j}{2} - \frac{3^{2/3}}{5} \epsilon^{2/3} + O(\epsilon) \right) \nonumber
\end{eqnarray}
where as before $\epsilon = 1-y$ and $j$ is the constant defined in \eqref{eq:constant}, and at late times goes as,
\begin{eqnarray}
{g}^{(1,p)}_{ij}(y,x) = \bar{S}^{(1)}_{ij}(x) y^2 + O(y^5) .\nonumber
\end{eqnarray}
The homogeneous solution is given by,
\begin{eqnarray}
{g}^{(1,h)}_{ij}(y,x)  = \bar{a}_{ij}(x) + \frac{ 3 y^3 }{1-y^3} \bar{b}_{ij}(x)  \nonumber
\end{eqnarray}
with constants of integration $\bar{a}_{ij}(x), \bar{b}_{ij}(x)$ which we may regard as tensors living on the geometry defined by $\bar{g}_{ij}$, and behaves as,
\begin{eqnarray}
{g}^{(1,h)}_{ij}(y,x)  = \bar{b}_{ij}(x) \frac{1}{\epsilon} + \left(  \bar{a}_{ij}(x)  - 2  \bar{b}_{ij}(x)  \right) + O( \epsilon ) \nonumber
\end{eqnarray}
at early times, and at late times goes as,
\begin{eqnarray}
{g}^{(1,h)}_{ij}(y,x)  = \bar{a}_{ij}(x) + 3 \bar{b}_{ij}(x)  y^3 + O(y^6).\nonumber
\end{eqnarray}

We must now impose boundary conditions to solve the system. As stated, at early times we wish to impose inflationary initial conditions, namely that the perturbation has no decaying behaviour in the early matter epoch. Hence we choose the homogeneous data $\bar{b}_{ij}(x)$ to cancel the $1 / \epsilon$ behaviour of the particular solution. The remaining data $\bar{a}_{ij}$ simply represents a change in the boundary metric $g_{ij}(0,x) = \bar{g}_{ij} + \bar{a}_{ij} + O(\lambda^4)$ at order $\lambda^2$. As discussed above, we are free to choose the boundary metric to be given by $\bar{g}_{ij}$ and not be perturbed at any order in the expansion. Hence we require $\bar{a}_{ij} = 0$. Thus we have chosen,
 \begin{eqnarray}
{g}^{(1)}_{ij}(y,x)  = k^{(1)}(y) \bar{S}^{(1)}_{ij}(x)  \nonumber
\end{eqnarray}
with time dependence,
 \begin{eqnarray}
k^{(1)}(y) = \left( - j \frac{ 3 y^3 }{1-y^3} +  \frac{y^2  \left( 2 + (1 + y^3)\; {}_2F_1\left[ 1,\frac{4}{3};\frac{5}{3}; y^3 \right] \right)}{3 (1 - y^3)^{1/3}} \right). \nonumber
\end{eqnarray}
Since the equations here are of the form in the scalar perturbation theory above, it is no coincidence that we have found the same time dependence as in \eqref{eq:scalarPTsoln}. 

The constraint equation at this order, $\Phi^{(1)}_i = 0$, is non-trivial, and yields,
\begin{eqnarray}
\Phi^{(1)}_i = \bar{\nabla}^j \dot{g}^{(1)}_{ij} -  \bar{\nabla}_i \dot{g}^{(1)}\nonumber
\end{eqnarray}
where indices are raised/lowered with respect to $\bar{g}_{ij}$.
Since $\bar{\nabla}^j \bar{S}^{(1)}_{ij} -  \bar{\nabla}_i \bar{S}^{(1)} = 0$, this implies the constraint is indeed satisfied by our leading order solution.

The dust density is also modified at leading order, and we can compute,
\begin{eqnarray}
\rho^{(1)}(y,x)  = - H^2 y^2 \left( \ddot{g}^{(1)} + \left( - \frac{1}{y} + 2 \frac{\dot{a}}{a} \right) \dot{g}^{(1)} \right)\nonumber
\end{eqnarray}
from the scalar equation, and hence,
\begin{eqnarray}
\rho^{(1)}(y,x) =  - \frac{y^2}{4} \left( \ddot{k}^{(1)} + \left( - \frac{1}{y} + 2 \frac{\dot{a}}{a} \right) \dot{k}^{(1)} \right)  \bar{R}(x). \nonumber
\end{eqnarray}
In particular, we find that the perturbation to the FLRW zeroth order density by the first order goes at early times as,
\begin{eqnarray}
\frac{\rho^{(1)}}{\rho^{(0)}} = \frac{3^{2/3}}{40 H^2} \bar{R}(x) \epsilon^{2/3} + O(\epsilon^{5/3})\nonumber
\end{eqnarray}
which has the same growing form as the in the linear calculation. Had we not chosen the growing mode initial condition, and therefore had a perturbation involving some decaying behaviour, we would have found $\rho^{(1)} / \rho^{(0)} \sim 1 / \epsilon$ at early times, and hence the universe would not have approached FLRW in the far past.
We see explicitly that for our choice of initial conditions the deformation $g^{(1)}_{ij}(y,x)$ is a small perturbation to the leading behaviour for all times, including during the early matter era, and hence this order of the gradient expansion gives a controlled approximation for all the $\Lambda$ and matter epochs.

\subsection{Solution at four derivative order: $O( \lambda^4 )$}

At four derivative order, $O(\lambda^4)$, the equations are a little more complicated, and are given as,
\begin{eqnarray}
D_y \partial_y {g}^{(2)}_{ij}(y,x) = S^{(2,a)}_{ij}(y,x) + S^{(2,b)}_{ij}(y,x)\nonumber
\end{eqnarray}
where $S^{(a)}$ and $S^{(b)}$ are sources for the same linear differential operator, $D$, as at lower order, and depend on the solution to the lower orders. Explicitly these are given as,
\begin{eqnarray}
S^{(2,a)}_{ij} & =&  \frac{1}{a^2 H^2} \Big[ \bar{\nabla}^2 g^{(1)}_{ij} - \bar{\nabla}_i \bar{\nabla}_j g^{(1)} - 2 \bar{\nabla}_{(i} v_{j)} + \frac{1}{2} \bar{g}_{ij}  \bar{\nabla}^m v_m \nonumber \\
&& \qquad + \frac{3}{2} \left( \bar{R} g^{(1)}_{ij} + \bar{g}_{ij} \bar{R}^{mn} g^{(1)}_{mn} - 4 \bar{R}_{(i}^{~m} g^{(1)}_{j) m} \right)
\nonumber \\
&& \qquad + 2 \left( \bar{R}_{ij} - \frac{1}{2} \bar{g}_{ij} \bar{R} \right) g^{(1)} \Big]
\nonumber \\
S^{(2,b)}_{ij} & = & \dot{g}^{(1)}_{im} \dot{g}^{(1)m}_{~j} - \frac{1}{2} \dot{g}^{(1)} \dot{g}^{(1)}_{ij} - \frac{1}{8} \left( \dot{g}^{(1)}_{mn} \dot{g}^{(1)mn} - (\dot{g}^{(1)})^2 \right) \bar{g}_{ij}\nonumber
\end{eqnarray}
where indices are raised/lowered with respect to $\bar{g}_{ij}$ and $v_i = \bar{\nabla}^m g^{(1)}_{mi} - \bar{\nabla}_i g^{(1)}$, which vanishes given the form of $g^{(1)}_{ij}$ given above.
This then yields the solution,
\begin{eqnarray}\label{gfour}
{g}^{(2)}_{ij}(y,x) = k^{(2,a)}(y) \bar{S}^{(2,a)}_{ij} + k^{(2,b)}(y) \bar{S}^{(2,b)}_{ij}\nonumber
\end{eqnarray}
where,
\begin{eqnarray}
\bar{S}^{(2,a)}_{ij} & =&  \frac{1}{H^4} \Big[ \bar{\nabla}^2 \bar{R}_{ij} - \frac{1}{4}\bar{\nabla}^2\bar{R} \bar{g}_{ij} -  \frac{1}{4}\bar{\nabla}_i\bar{\nabla}_j \bar{R}\nonumber\\
&& - \bar{R}^2 \bar{g}_{ij} +\frac{7}{2} \bar{R}\bar{R}_{ij} + \frac{3}{2} \bar{R}^{mn}\bar{R}_{mn}\bar{g}_{ij} - 6 \bar{R}_i^{~m}\bar{R}_{mj}\Big]
\nonumber \\
\bar{S}^{(2,b)}_{ij} & = &  \frac{1}{H^4} \Big[ \frac{9}{64}\bar{R}^2 \bar{g}_{ij} -\frac{5}{8} \bar{R}\bar{R}_{ij} -\frac{1}{8} \bar{R}^{mn}\bar{R}_{mn}\bar{g}_{ij}\nonumber \\
&&+\bar{R}_i^{~m}\bar{R}_{mj}\Big]\nonumber
\end{eqnarray}
and the functions $k^{(2,a/b)}$ solve,
\begin{eqnarray}
D_y \partial_y k^{(2,a)} =  \frac{k^{(1)}(y)}{a^2(y)} \; , \quad D_y \partial_y k^{(2,b)} =  ( \dot{k}^{(1)}(y) )^2\quad \label{einstein2nd}
\end{eqnarray}
such that they have no decaying behaviour at early times, so $k^{(2,a/b)} \sim O( \epsilon^0 )$ as $\epsilon \rightarrow 0$, and they do not lead to a deformation of $\bar{g}_{ij}$, so that at late times $k^{(2,a/b)} \sim O( y^3 )$ for $y \rightarrow 0$. Note that unlike $O(\lambda^2)$ the late time fall off goes as $O(y^3)$ rather than $O(y^2)$. This is simply due to the fact that we know the $O(y^2)$ late time behaviour from the late time expansion \eqref{eq:soln} only involves 2 derivatives. Hence order $O(\lambda^4)$ and higher cannot contribute to it, provided the boundary metric is fixed to be $\bar{g}_{ij}$.

\begin{figure}[h!]
\includegraphics[width=0.4\textwidth]{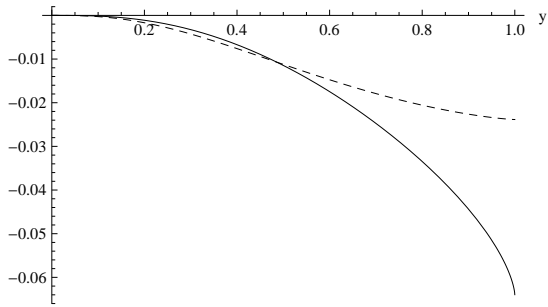}
\caption{\label{figk}The functions $k^{(2,a)}(y)$ (solid) and $k^{(2,b)}(y)$ (dashed).}
\end{figure}

Whilst we cannot compute the functions $k^{(2,a/b)}$ analytically, numerically they are simple to determine. We give plots of the functions in figure \ref{figk}. 
They have the following behaviour at late times,
\begin{eqnarray}\label{latetimek}
k^{(2,a/b)} & = &   c_{a/b} y^3 + O(y^4)
\end{eqnarray}
where $c_a$ and $c_b$ will shortly be given numerically. However we can derive a relation between them. Combining the equations in \eqref{einstein2nd} we find,
\begin{equation}
 D_y\left(\dot{k}^{(2,b)}-2\dot{k}^{(2,a)}-k^{(1)}\dot{k}^{(1)}\right)=0,\label{subtractedeinstein}
 \end{equation}
with a solution $\dot{k}^{(2,b)}-2\dot{k}^{(2,a)}-k^{(1)}\dot{k}^{(1)}=0$, to which we may add the homogeneous mode of $D_y$. Since we require that $k^{(2,a)}$ and $k^{(2,b)}$ do not diverge at early times, and vanish at late times, we find that no addition of homogeneous mode to this particular solution is allowed. Expanding this solution at late times we find,
\begin{equation}
c_b=2c_a.\label{cabrelation}
\end{equation}
We find numerically
\begin{equation}\label{eqca}
c_a\simeq -0.184
\end{equation}
and confirm to the same precision that the relation \eqref{cabrelation} holds. As a further check, we may study the linearisation of equation \eqref{gfour}. In a late time expansion, the coefficient of $y^3$ agrees with that obtained numerically in the tensor-perturbation calculation, \eqref{eqc2}, since $c_a = -2 c_2$ to the precision obtained.

The constraint equation at this order, $\Phi^{(2)}_i = 0$, gives,
\begin{eqnarray}
\label{eq:constsecondorder}
\Phi^{(2)}_i &=& \bar{\nabla}^j \dot{g}^{(2)}_{ij} -  \bar{\nabla}_i \dot{g}^{(2)} - \bar{\nabla}_m ( g^{(1)mn} \dot{g}^{(1)}_{ni} )\nonumber\\
& &  + \bar{\nabla}_i ( g^{(1)mn} \dot{g}^{(1)}_{mn} )+ \frac{1}{2} (\bar{\nabla}^m g^{(1)} ) \dot{g}^{(1)}_{mi}\nonumber\\
&& - \frac{1}{2} ( \bar{\nabla}_i g^{(1)}_{mn} ) \dot{g}^{(1)mn}
\end{eqnarray}
with indices raised/lowered with $\bar{g}_{ij}$,
and must vanish for a solution. On our solution above, this translates into the condition,
\begin{equation}
\dot{k}^{(2,b)}-2\dot{k}^{(2,a)}-k^{(1)}\dot{k}^{(1)}=0\nonumber
\end{equation}
which as discussed previously is satisfied because it is the solution to equation \eqref{subtractedeinstein} which is consistent with our chosen boundary conditions.
The density at this order behaves as,
\begin{eqnarray}
\rho^{(2)}_i(y,x)  &=& - H^2 y^2 \Bigg( \ddot{g}^{(2)} + \left( - \frac{1}{y} + 2 \frac{\dot{a}}{a} \right) \dot{g}^{(2)} - \frac{1}{2} \dot{g}^{(1)ij} \dot{g}^{(1)}_{ij} \nonumber \\
&&  - g^{(1)ij} \left( \ddot{g}^{(1)}_{ij} + \left( - \frac{1}{y} + 2 \frac{\dot{a}}{a} \right) \dot{g}^{(1)}_{ij} \right) \Bigg)\label{rho2}\nonumber
\end{eqnarray}
and in fact we can explicitly compute the leading order early time behaviour analytically as,
\begin{eqnarray}
\frac{\rho^{(2)}}{\rho^{(0)}} = -\frac{3^{5/3}}{160} \frac{j}{H^4} \left(\bar{R}_{mn}\bar{R}^{mn} - \frac{1}{2}\bar{R}^2\right) \epsilon^{2/3} + O(\epsilon^{5/3}).\nonumber
\end{eqnarray}
Again we see that for our choice of inflationary initial conditions this order of the gradient expansion gives a controlled approximation for the full duration of the $\Lambda$ and matter eras.

\subsection{Solution at general order}

We believe the structure of the tensor equation at $O(\lambda^{2n})$  order for $n>4$ will be,
\begin{eqnarray}
D_y \partial_y {g}^{(n)}_{ij}(y,x) = \sum_a p^{(n,a)}(y) \bar{S}^{(n,a)}(x) \nonumber
\end{eqnarray}
where the sum of terms on the right hand side is a source for the linear problem with operator $D$. The spatial dependence of the source terms enters in the expressions $\bar{S}^{(n,a)}$ which are composed from the solutions of the lower orders, $\bar{S}^{(m,b)}$ for $m < n$, by tensor contractions, constractions with $\bar{R}_{ij}
$, and differentiation by $\bar{\nabla}$. Likewise the $y$ dependence of the source is given by the functions $p^{(n,a)}$ which are composed of expressions involving the $p^{(m, b)}$ for $m<n$, the scale factor, and differentiation $\partial / \partial y$.

The required solution of this tensor equation is then,
\begin{eqnarray}
{g}^{(n)}_{ij}(y,x) = \sum_a q^{(n,a)}(y) \bar{S}^{(n,a)}(x)\nonumber
\end{eqnarray}
where the functions $q^{(n,a)}$ solve,
\begin{eqnarray}
D_y \partial_y q^{(n,a)} = p^{(n, a)} \nonumber
\end{eqnarray}
and the homogeneous modes are chosen so that $q^{(n,a)} \sim O(y^3)$ at late times (so the boundary metric $\bar{g}_{ij}$ is not perturbed), and there is no decaying behaviour $\sim 1/\epsilon$ at early times as $\epsilon \rightarrow 0$, so we expect $q^{(n,a)} \sim O(\epsilon^0)$, ensuring the gradient expansion gives a controlled approximation at early times.

A subtle point is whether such a solution obeys the constraint $\Phi^{(n)}_i = 0$. We have seen that at orders $n = 1,2$ this is satisfied, but it is not immediately obvious that we should have expected this, and we have seen that at order $n=2$ the constraint is satisfied in a non-trivial way. We now provide an argument that for our boundary conditions the constraint should always hold.

Let us assume that we have satisfied all the constraints up to order $n-1$, so $ \Phi^{(m)}_i(y,x) = 0$ for all $m < n$, and we have found the $n$-th order solution $g^{(n)}_{ij}$ to the tensor equations, and that as claimed above, at early times the solutions behaves as $ g^{(m)}_{ij} \sim O(\epsilon^0)$ for all $m \le n$ since we have selected inflationary boundary conditions at each order. Now consider the equation \eqref{eq:dconst}. On the right hand side $\dot{g} \sim O( \lambda^2 )$, and hence if  we consider order $O(\lambda^{ 2 n+1})$ for this equation, we find the simple condition,
\begin{eqnarray}
D_y \Phi^{(n)}_i = 0 \nonumber
\end{eqnarray}
which implies $\Phi^{(n)}_i$ must behave as,
\begin{eqnarray}
\Phi^{(n)}_i(y,x) & = & b_i(x) \frac{y^2}{(1 - y^3)^2} \nonumber
\end{eqnarray}
for some constants of integration $b_i(x)$.
Now $\Phi^{(n)}_i$ depends on both $g^{(n)}_{ij}$ and all previous orders $g^{(m)}_{ij}$ with $m < n$ (for example, see equation \eqref{eq:constsecondorder}). The form of the constraint equation implies that provided $g^{(m)}_{ij} \sim O(\epsilon^0)$ for all $m \le n$ at early times, then $\Phi^{(n)}_i$ should remain finite as $\epsilon \rightarrow 0$, since there are no explicit divergences in the constraint equation, and so a divergence in $\Phi^{(n)}_i$ could only arise from a divergence in $g^{(m)}_{ij}(y,x)$ for some $m \le n$. However, we see from above that for $b_i(x) \ne 0$ the constraint $\Phi^{(n)}_i \sim O(1 / \epsilon^{2})$ at early times. Hence we deduce that for our inflationary boundary conditions we must have $b_i(x) = 0$ and thus the constraint is satisfied. It is interesting that the constraint is not automatically satisfied, and it is an artifact of our inflationary choice of boundary condition that ensures this. We may regard the fact that the vector constraint equation gives no obstruction to the moduli approximation at arbitrary order as confirmation that our choice of normal coordinates to the future boundary, together with the choice that these comove with the dust, is indeed consistent.

\section{Boundary stress tensor}
\label{sec:stress}

The linear theory led us to conjecture that cosmologies which in the past tend to FLRW, in the sense that the initial perturbation is a growing one, are characterized only by the metric $\bar{g}_{ij}$, and the tensor $h_{ij}$ is completely determined in terms of it. Using our moduli construction we have explicitly constructed such cosmologies as a gradient expansion up to 4 derivative order. Calculating the boundary tensor $\bar{h}_{ij}$ to this order yields,
\begin{eqnarray}\label{hresult}
\bar{h}_{ij} &=& - \frac{4}{3} \bar{g}_{ij} - \frac{3 j}{H^2} \left(  \bar{R}_{ij} - \frac{1}{4} \bar{g}_{ij} \bar{R} \right) \nonumber  \\
&& + \frac{c_a}{H^4} \Big[ \bar{\nabla}^2 \bar{R}_{ij} - \frac{1}{4}\bar{\nabla}^2\bar{R} \, \bar{g}_{ij} -  \frac{1}{4}\bar{\nabla}_i\bar{\nabla}_j \bar{R} \nonumber \\
&& - 4 \bar{R}_i^{~m}\bar{R}_{mj} + \frac{5}{4} \bar{R}^{mn}\bar{R}_{mn}\bar{g}_{ij} +\frac{9}{4} \bar{R}\bar{R}_{ij} - \frac{23}{32} \bar{R}^2 \bar{g}_{ij}  \Big]  \nonumber \\
&& + O( \lambda^6 )\nonumber
\end{eqnarray}
where $c_a$ is defined in \eqref{latetimek} and given numerically in \eqref{eqca}.  From this, a corresponding expression for the boundary stress tensor may be obtained, $\bar{T}_{ij} = \bar{h}_{ij} - \bar{g}_{ij} \bar{h}$. We note that this is indeed of the form predicted by the linear calculation in equation \eqref{eq:predict}. We note that the 4 derivative term above does indeed vanish on a scalar perturbation, as required for consistency with the linear analysis. From a holographic perspective we would say that the inflationary initial condition corresponds to the dual QFT being in a particular state, associated to the metric $\bar{g}_{ij}$ on which it is defined.

We may also re-express the tensor $\bar{T}_{ij}$ by generating it through the variation of a Euclidean action, $S[\bar{g}^{ij}]$,
\begin{equation}
\bar{T}_{ij} \equiv \frac{1}{\sqrt{\bar{g}}} \frac{\delta S[\bar{g}^{ij}]}{\delta \bar{g}^{ij}}. \nonumber
\end{equation}
Generating $\bar{T}_{ij}$ in this way guarantees that it is automatically transverse, provided $S[\bar{g}^{ij}]$ is diffeomorphism invariant.
We find the action which yields the derivatively expanded $h_{ij}$ in \eqref{hresult} is given by,
\begin{eqnarray}
S[\bar{g}^{ij}] &=&  \int d^3x \sqrt{\bar{g}}\Bigg[  \frac{8}{3}-\frac{3j}{H^2}\bar{R} \nonumber\\
&&+\frac{c_a}{H^4}\left(\bar{R}_{mn}\bar{R}^{mn} - \frac{3}{8} \bar{R}^2\right)+ O( \lambda^6 )\Bigg].\nonumber
\end{eqnarray}

In the case of AdS-CFT, there has been considerable study of the moduli space deformations about the flat homogeneous AdS-Schwarzschild solution \cite{Bhattacharyya:2008jc,Bhattacharyya:2008xc,Baier:2007ix}. There, requiring the conformal boundary metric to be unperturbed and the horizon to be regular, one finds that the deformed solutions are characterized by the solutions of a dual fluid dynamics. We note that in this case, even for a trivial boundary metric there are interesting configurations of the boundary stress tensor. By contrast, in our cosmological setting, the stress tensor and solutions become trivial for a trivial boundary metric $\bar{g}_{ij} = \delta_{ij}$. We believe this is a reflection of the fact that the dual holographic theory is Euclidean rather than Lorentzian, and hence would be expected to have a more restricted dynamics.

We note that in the Hamilton-Jacobi formalism one also obtains a functional for a 3-metric \cite{PhysRevD.42.3936,Salopek:1990re,Parry:1993mw}, and indeed at fourth order in a derivative expansion, the same combination of curvature invariants is found \cite{Parry:1993mw}. It would be interesting to explore this connection further.

\section{Smoothing post-inflationary $\Lambda$-dust cosmologies}
\label{sec:smoothing}

Since a post-inflationary $\Lambda$-dust cosmology is characterized by a Euclidean 3-metric, it is rather natural to consider flows in the space of solutions.
A canonical example of such a flow is that given by the 3-dimensional Ricci flow, $\partial \bar{g}_{ij} / \partial \tau = - 2 \bar{R}_{ij}$, where $\tau$ is the auxiliary flow time (unrelated to real time). Since this acts only on the data $\bar{g}_{ij}$ which characterizes our post-inflationary $\Lambda$-dust cosmologies then by construction it gives a flow in the space of these cosmologies. Furthermore, it has three interesting properties. Firstly it provides a well posed parabolic flow. Secondly,  we may interpret the flow from a starting 3-metric as a short distance smoothing procedure, where a flow for a time $\tau_0$ will result in a smoothing of a range of scales in the boundary metric (and hence cosmology) with characteristic scales shorter than $\sim 1/\sqrt{\tau_0}$. Thirdly the flat FLRW solution, $\bar{g}_{ij} = \delta_{ij}$ is a (linearly stable) fixed point of the flow.  

The use of Ricci flow for averaging cosmological data is also considered by Buchert and Carfora\cite{Buchert:2002ht,Buchert:2003,Buchert:2008zm}. However, an essential difference here is that for our solutions the data associated to the stress tensor is determined by the 3-metric -- a consequence of imposing our inflationary initial conditions. Thus we need only to smooth the 3-metric data and not the extrinsic curvature too.

\section{Summary}
\label{sec:summary}

As an approximation to describing the $\Lambda$ and matter eras of our $\Lambda$CDM universe we have solved the $\Lambda$-dust Einstein equations using a moduli space expansion and impose initial conditions consistent with inflation, namely that in the early matter era, the cosmology tends to a flat FLRW solution. Our solutions are fully non-linear in the amplitude of the metric deformation away from FLRW. Furthermore, precisely because of our choice of initial condition they yield a controlled approximation from the earliest times in the matter epoch right through to the far $\Lambda$ dominated future.

An important simplifying step was the adoption of comoving coordinates \eqref{eq:metric1}, for which the moduli are the components of the comoving FLRW spatial three-metric. This data is then promoted to vary on the comoving slices, and consequently the Einstein equations are no longer satisfied. Solutions may then obtained by introducing time-dependent correction terms to the spatial three-metric and the dust density, order by order in spatial derivatives. Following this procedure we explicitly constructed solutions up to fourth order in comoving derivatives, imposing the inflationary initial conditions. We have also described the structure of the solution we expect at general order, and in particular we have demonstrated that the vector constraint equation gives no obstruction. Our construction appears to reply on the presence of the cosmological constant today which controls all scales in our solutions. More generally the unnatural smallness of this late time scale allows a decoupling of the details of the early universe, such as the inflation and radiation era, from the dynamics of the late time universe, the matter and $\Lambda$ eras.
As remarked in the introduction, the moduli space approximation we use here is strongly reminiscent to the derivation of hydrodynamics from gravitational solutions in AdS/CFT.

In a general late time characterisation of $\Lambda$CDM solutions, as discussed in \cite{Wiseman:2010kp}, a $\Lambda$-dust universe dominated by $\Lambda$ in the future is determined by data on its future conformal boundary, namely a 3-metric and stress tensor. We have found that our choice of inflationary initial conditions determines the stress tensor in terms of the 3-metric. Whilst such a reduction in data is to be expected when one moves from a general solution to one obeying certain initial conditions, it is perhaps a little surprising that the expression for the stress tensor is a \emph{local} one in terms of the 3-metric.

Whilst one might have thought that characterizing our cosmology by using `future' data rather than initial data would be practically hopeless if one was interested in particular initial conditions, such as inflationary ones, we have seen that at least on large scales this is not the case at all. It then remains an interesting question
whether we may use the construction presented here to sharpen large scale observational constraints at late times using late time observations, for example using the luminosity-distance vs. redshift relation computed in \cite{Wiseman:2010kp}, even when we make the assumption of inflationary initial conditions.

\section*{Acknowledgements}

We would like to thank Carlo Contaldi, Carsten Gundlach, Shinji Mukohyama, Arttu Rajantie, Tetsuya Shiromizu, Takahiro Tanaka, Paul Tod and James Vickers for useful and stimulating discussions on this topic. TW is supported by an STFC advanced fellowship and Halliday award. BW was supported by an STFC studentship and is currently supported by a Royal Commission for the Exhibition of 1851 Research Fellowship.

\bibliographystyle{apsrev}
\bibliography{moduli}

\end{document}